\documentstyle[epsbox,subfigure]{mn}
\def\lsim{\mathrel{\mathpalette\Oversim<}}
\def\gsim{\mathrel{\mathpalette\Oversim>}}
\def\Oversim#1#2{\lower0.5ex\vbox{\baselineskip0pt\lineskip0pt%
            \lineskiplimit0pt\ialign{%
          $\mathsurround0pt #1\hfil##\hfil$\crcr#2\crcr\sim\crcr}}}

\newcommand{\apj}{ApJ}
\newcommand{\apjs}{ApJS}
\newcommand{\mnras}{MNRAS}
\newcommand{\aap}{A\&A}
\def\goodgap{
\hspace{\subfigtopskip}
}
\title{UV Background-Induced Bifurcation of the Galactic Morphology}
\author[Susa \& Umemura]{Hajime~Susa \thanks{e-mail:susa@rccp.tsukuba.ac.jp}
and Masayuki~Umemura \thanks{e-mail:umemura@rccp.tsukuba.ac.jp}\\
Center for Computational Physics, University of
  Tsukuba, Tsukuba 305, Japan }
\date{Accepted 2000 May}
\pubyear{2000}
\begin{document}
\maketitle
%%%%%%%%%%%%%%%%%%%%%%%%%%%%%%%%%%%
% (2)Abstract  & Subject Headings %
%%%%%%%%%%%%%%%%%%%%%%%%%%%%%%%%%%%

\begin{abstract}
Based upon a novel paradigm of the galaxy formation under UV background,
the evolutionary bifurcation of pregalactic clouds is 
confronted with observations on elliptical and spiral galaxies.
The theory predicts that the dichotomy between 
the dissipational and dissipationless galaxy formation
stems from the degree of self-shielding from the UV background 
radiation. This is demonstrated on a bifurcation diagram of
collapse epochs versus masses of pregalactic clouds.
Using the observed properties, the collapse epochs are 
assessed for each type of galaxies with attentive mass estimation. 
By the direct comparison of the theory with the observations, 
it turns out that  
the theoretical bifurcation branch successfully discriminates
between elliptical and spiral galaxies.
This suggests that the UV background radiation 
could play a profound role for the differentiation 
of the galactic morphology into the Hubble sequence.
\end{abstract}
\begin{keywords}
galaxies: formation --- radiative transfer --- --- molecular processes
\end{keywords}

%%%%%%%%%%%%%%%%%%%%%%%%%%%%%
% (3)TEXT & Acknowledgments %
%%%%%%%%%%%%%%%%%%%%%%%%%%%%%
\section{Introduction}
\label{intro}
A substantial basis of the galaxy formation theory has been founded 
by several pioneering works in 1970s (\cite{RO77}; \cite{Silk77}). 
The theory predicts that the galactic scales are basically
determined by atomic cooling of hydrogen and partially helium.
The pregalactic evolution is elegantly summarized on the {\it cooling diagram} 
for virialized objects.    
%%%%%
Moreover, the origin of the Hubble type has been attributed to
the dissipativeness of the collapse, which is regulated by the efficiency
of star formation. If the star formation
proceeds after most of the gravitational energy 
is dissipated, the pregalactic clouds will evolve into spiral galaxies. 
This is a paradigm of the so-called {\it dissipational galaxy formation}
(e.g. \cite{Lar76}; \cite{Car85}; \cite{KG91}).
On the other hand, an early star formation episode 
leads to the {\it dissipationless galaxy formation}
(e.g. \cite{AB78}; \cite{AM90}),
ending up with the formation of elliptical galaxies.
However, the key mechanism which physically controls the star formation
efficiency has been hitherto unsolved. 

Very recently, Susa \& Umemura (2000) 
(hereafter SU) propose that
the self-shielding against UV background radiation regulates the
star formation in pregalactic clouds.
The star formation processes in primordial gas
have been explored by many authors with {\it ab initio} 
calculations (e.g. \cite{MST69}; \cite{Hut76}; \cite{Car81};
\cite{PSS83}; \cite{SUN96}; \cite{Uehara96}; \cite{ON98} 
; \cite{Nishi98}; \cite{NU99}). 
The key physics is the radiative cooling by H$_2$ line emission, 
because H$_2$ is the only coolant for primordial gas 
at $T\lsim 10^4$ K. 
SU have studied the efficiency of H$_2$ cooling 
in collapsing clouds exposed to UV background radiation,
because it is significant after the reionization of the universe, 
probably at $z\lsim 10$ (\cite{NUS99}, and references therein).
SU have found that if a cloud undergoes the first sheet collapse 
at higher redshifts ($z\gsim 4$),
then the cloud is quickly shielded against the UV background and
consequently cools down due to the efficient formation of H$_2$.
Resultantly, it leads to an early burst of star formation.
On the other hand, the shielding is retarded for a later collapsing cloud
at $z\lsim 4$, resulting in the dissipational galaxy formation. 
As a result, the bifurcation branch of the self-shielding 
is corresponding to the
boundary between the dissipationless and dissipational galaxy formation. 
In this {\it Letter}, this novel bifurcation theory 
is confronted with the observations of elliptical and spiral galaxies to
elucidate whether the theory is practically successful or not.
 
\section{Bifurcation Theory}
\label{BITH}
In Figure 1, we show the bifurcation theory. 
Here, the cosmological parameters are assumed to be $\Omega=0.3$,
$h=0.7$, and $\Omega_{\rm b} h^2=0.02$
with usual meanings. 
Originally, 
the sheet collapse was pursued with two initial parameters, i.e., 
the mean density ($\bar{n}_{\rm ini}$) and the thickness ($\lambda$).
Here, the initial parameter space is translated into the baryonic mass 
[$M_{\rm b}\equiv (4\pi/3) \bar{n}_{\rm ini}(\lambda/2)^3$]
and the collapse epoch ($z_{\rm c}$) by 
assuming the initial stage is close to the maximum expansion
of a density fluctuation.
In the region (a) in Figure 1, a pregalactic cloud is self-shielded
against the external UV in the course of the sheet collapse, so that
the cloud cools down below $10^3$K and undergoes efficient star formation.
Hence, it is expected to evolve into an early type galaxy with 
a large bulge-to-disk ratio (B/D) due to the dissipationless virialization.
In the region (b), the cloud is not self-shielded during the sheet 
collapse, but will be self-shielded through the shrink to the rotation
barrier. This leads to the retarded star formation, and thus
the virialization would proceed in a fairly dissipative fashion. 
As a result, a late type
(small B/D) galaxy would be preferentially born. The region (c)
represents the forbidden region of the collapse due to the Jeans 
stability.

SU have considered only baryonic component, because, 
in the later collapsing phase, the cooling sheet 
is dominated by baryons, not by the
diffuse dark matter. Then, if the UV background is constant,
the bifurcation mass scale is given as
\begin{eqnarray}
M_{\rm SB}^{\rm max}=2.2\times 10^{11} M_\odot 
\left(\frac{1+z_{\rm c}}{5}\right)^{-4.2}
\left(\frac{I_{21}}{0.5}\right)^{0.6} \label{eq:msb},
\end{eqnarray}
where $I_{21}$ is the UV background intensity in units of
$10^{-21}{\rm erg~ s^{-1} cm^{-2} str^{-1} Hz^{-1}}$.
%%%%%%%%%%%%%%%%%%%
However, the dark matter potential may affect the
evolution in the early phase of collapse. 
If we take this effect into account maximally, 
the Jeans length is reduced by a factor of 
$\sqrt{\Omega_{\rm b}/\Omega}$. 
With this Jeans scale,
the bifurcation mass is changed to be
$M_{\rm SB}^{\rm min}=\left(\Omega_{\rm b}/\Omega\right)^{1.2}
M_{\rm SB}^{\rm max}$. Thus, the practical bifurcation mass would be 
between $M_{\rm SB}^{\rm max}$ and $M_{\rm SB}^{\rm min}$.

\begin{figure}
\begin{center}
\subfigure[]{\psbox[width=80mm,vscale=1.0]{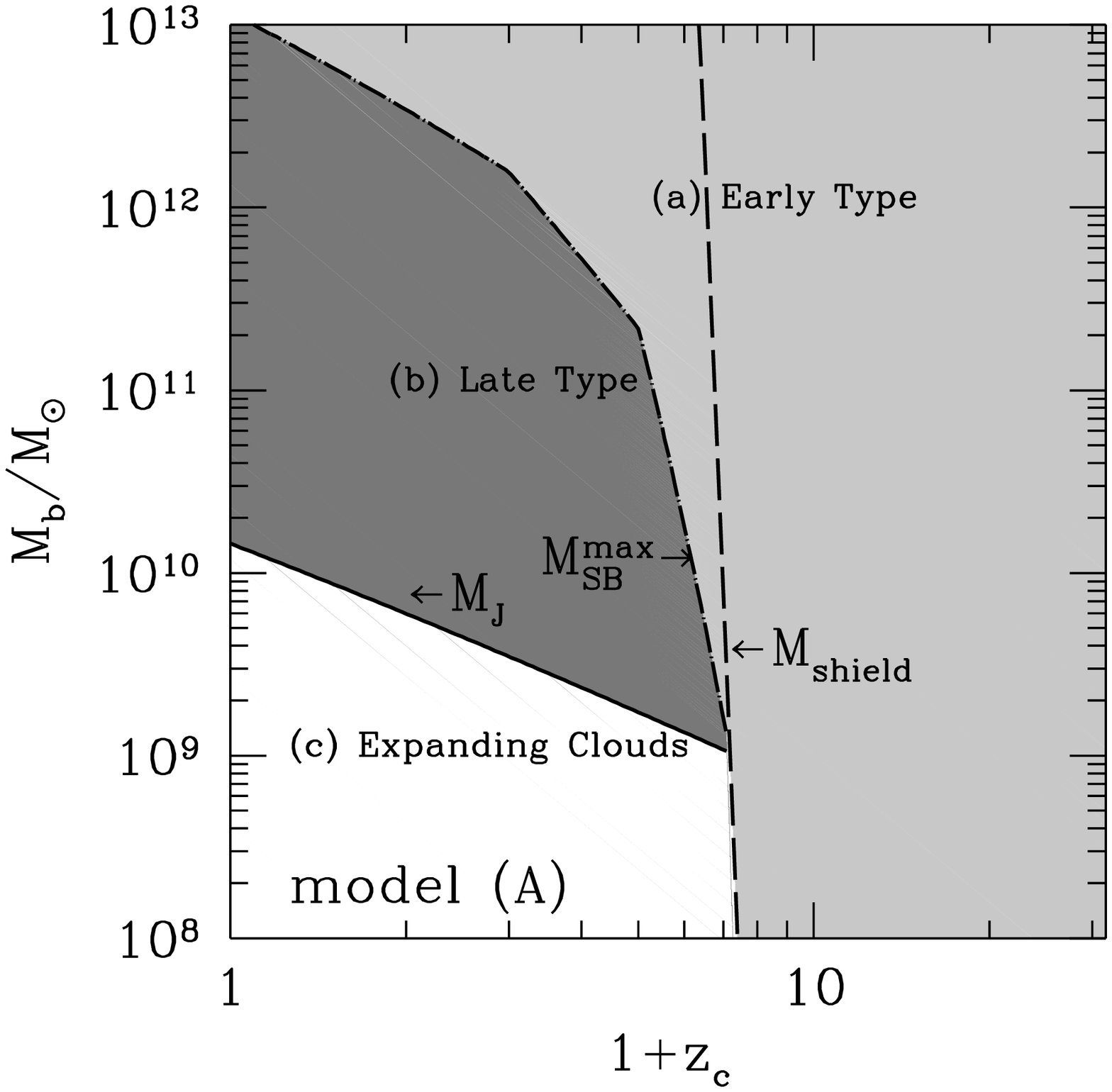}}\goodgap
\subfigure[]{\psbox[width=80mm,vscale=1.0]{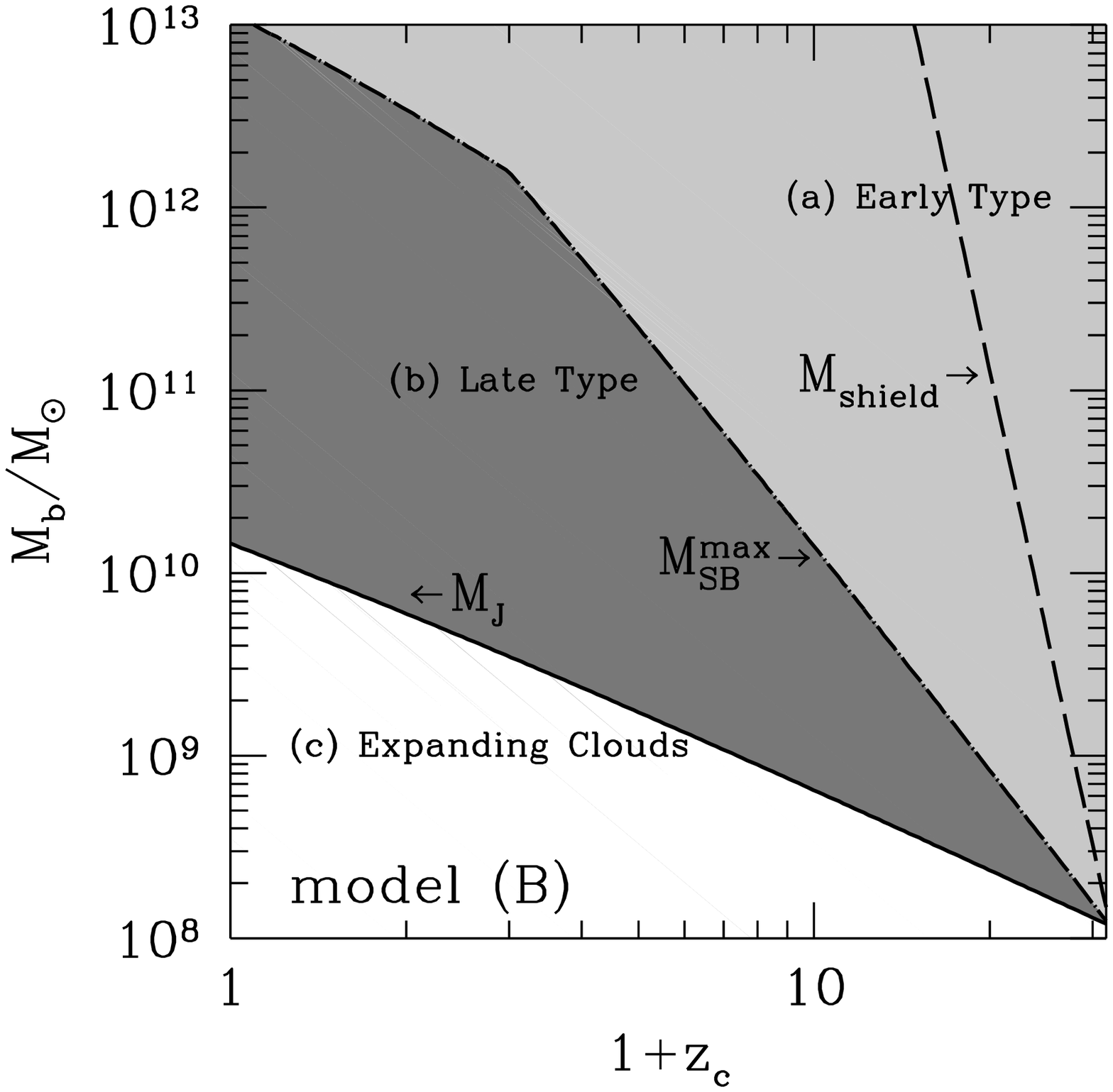}}
\end{center}
\caption[f1a.eps,f1b.eps]{The bifurcation diagram. The prediction of galactic 
evolution is illustrated on the plane of 
the baryonic mass ($M_{\rm b}$) versus collapse redshift ($z_c$).  
Fig.\ref{fig1}a corresponds to
the UV evolution of model A (see text), and Fig.\ref{fig1}b 
for model B (see also text). 
In both of the figures, the half-tone shaded area (a) 
denotes the region in which pregalactic clouds evolve into early type
galaxies. 
The shaded area (b) leads to the formation of late-type galaxies.
The unshaded area (c) represents the region in which
the density perturbations cannot grow into bound objects, due to the thermal
pressure of ionized gas. 
Here we employ $M_{\rm SB}^{\rm max}$ as the bifurcation mass,
although the effect of dark matter may reduce the mass by factor
$(\Omega_{\rm b}/\Omega)^{1.2}$ at most. 
$M_{\rm shield}$ denotes the mass 
above which the cloud is initially self-shielded. }
\label{fig1}
\end{figure}

Also, SU have assumed the UV intensity to be independent of time. 
Practically, the intensity seems to evolve.
Here we include the effect of the evolution of 
UV background radiation.
We assume $I_{21}=0.5\left[\left(1+z\right)/3 \right]^3$ for $z\le 2$ 
and $I_{21} =0.5$ for $2 < z \le 4$.
This dependence is consistent with the UV intensity
in the present epoch (\cite{Mal93}; \cite{DS94}), and the value inferred from 
the QSO proximity effects at high redshifts 
(\cite{BDO88}; 
Giallongo et al. 1996).
As for $z > 4$, two extreme models are employed. 
The first one is (A) the exponentially damping model,
$I_{21}=0.5\exp\left[3\left(4-z\right)\right]$
(\cite{Ume00}), and 
the second one is (B) the constant extrapolation model,
$I_{21}=0.5$. 
In Figure \ref{fig1}, these two extreme models are shown.
The difference emerges especially at $z_{\rm c} \gsim 7$. 

\section{Estimation for Galactic Masses and Collapse Epochs}
\label{INOBDA}
Here, based upon the observational data, we attempt to 
assess the total baryonic masses of ellipticals as well 
as spirals and their collapse epochs. Then, the observed galaxies
are compared with the theory in the bifurcation diagram.

To begin with, we evaluate the baryonic masses from B-band luminosities 
for ellipticals and from I-band luminosities for spirals
with the mass-to-light ratio; 
$
M_{\rm b}=\left[M_*/L\right]L/f_*, 
$
where $M_*$ is the total stellar mass and 
$f_*$ is the mass fraction of stellar component 
in the total baryonic mass.
The mass-to-light ratios $M_*/L$ are obtained theoretically as well as 
observationally.  
Based on the chemical evolution theory 
(\cite{KA97}),
the $M_*/L$ at B-band for ellipticals is $4-9$.
This value is consistent with the observational estimates 
of $4-7$
(\cite{Ber93}; \cite{Piz97}). 
In this paper, we adopt the luminosity-dependent values given 
by Kodama \& Arimoto (1997).
The $M_*/L$ at I-band for spirals is 
3.37(Sa), 2.91(Sb), 1.79(Sc), and 1.33(Sd) 
based on the code developed by Kodama
\& Arimoto (1997) with the S1 model in Arimoto, Yoshii,
\& Takahara (1992).
These are also consistent with the observed data 
in Rubin et al.(1985).
$f_*$ is theoretically calculated 
by the population synthesis model to be 
0.963(Sa), 0.908(Sb), 0.462(Sc), and 0.125(Sd), while
basically $f_*=1$ for ellipticals.
With the estimated baryonic masses, we can assess the total masses as
$M_{\rm tot}=M_{\rm b}(\Omega/\Omega_{\rm b})$.

Next, we estimate the collapse epochs with the help of virial theorem.
For the purpose, we use the observed 1-D velocity dispersion. 
We assume that the system is spherical and the dark halo has
isothermal distributions after the virialization. 
We suppose a density perturbation as 
$\delta_i (r) = \bar{\delta_i} g(r)$, where
$r$ is the comoving radial coordinate and
$g(r)$ is a function which satisfies the normalization 
as $\frac{3}{R^3}\int_0^R g(r)r^2dr = 1$,
with $R$ being the comoving radius of the perturbed region.
$\bar{\delta_i}$ is the spatially averaged $\delta_i(r)$
in the volume $r \le R$.
Summing up the initial kinetic energy and gravitational energy,
we have the initial total energy of the perturbed region as
\begin{eqnarray}
E_{\rm ini}=-\frac{3GM_{\rm tot}^2}{5R} 
\left(1+z_i\right)\bar{\delta_i}\left(1+\frac{2}{3}\phi \right)F,
\label{eqn:Eini}
\end{eqnarray}
where $z_i$ is the initial redshift, $\phi$ is the contribution from 
the peculiar velocities, 
and $M_{\rm tot}$ is the total mass enclosed 
within the comoving radius $R$. 
The factor $F$ is defined as 
\begin{eqnarray}
F &\equiv& \frac{5}{R^5}\int_0^R r^4 \bar{g}(r)dr, \label{eqn:defG2}
\end{eqnarray}
where $\bar{g}(r)\equiv \frac{3}{r^3}\int_0^r r'^2 g(r')dr'$.
In the linear regime, $g(r)\propto r^{1/(n+3)}$ around a density peak
(\cite{HS85}), where $n$ denotes the index of the CDM power
spectrum. Then  equation (\ref{eqn:defG2})
is readily integrated to give $F = 5/(2-n)$. 
We assume $n=-1$ throughout this paper, because it is the case
for galactic scales in CDM cosmologies.
The total energy of an observed galaxy is derived
in terms of the virial theorem;
$
E_{\rm obs}=-\frac{3}{2}M_{\rm tot}\sigma_{\rm 1D}^2,
$
where $\sigma_{\rm 1D}$ is the 1-D internal velocity dispersion 
of the galaxy. 
Assuming the energy conservation, i.e., $E_{\rm int}=E_{\rm obs}$,
we have
\begin{eqnarray}
\frac{\left(1+z_i\right)\delta_i\left(1+2\phi/3 \right)}{R} 
= \frac{5\sigma_{\rm 1D}^2}{2GM_{\rm tot}} F^{-1}.
\label{eqn:Econ}
\end{eqnarray}

On the other hand, the time when the outer boundary 
of a perturbation collapses is analytically obtained (\cite{Pee93})
as
\begin{eqnarray}
t=\frac{\pi}{\left(GM_{\rm tot}\right)^{1/2}}
\left(
\frac{R}{2 \left(1+z_i\right) \delta_i \left(1+2\phi/3 \right) 
}\right)^{3/2}.
\label{eqn:tff}
\end{eqnarray}
Finally, using equations (\ref{eqn:Econ}) and (\ref{eqn:tff}), 
we have the collapse epoch as
$
t=\pi GM_{\rm tot} F^{3/2}/5^{3/2} \sigma_{\rm 1D}^3. 
$
By equating this time with the Hubble time $t_{\rm H}(z_{\rm c})$,
$t$ can be translated into the collapse redshift, $z_{\rm c}$. 
Thus, it turns out that the velocity dispersion
is a key quantity to determine $z_{\rm c}$.
We use the observed stellar velocity dispersion 
as $\sigma_{\rm 1D}$ for elliptical galaxies. 
Also, for some luminous ellipticals, $\sigma_{\rm 1D}$ is estimated
by the X-ray temperature $T_X$ as
$\sigma_{\rm 1D} = \sqrt{kT_X/\mu m_p}$. The estimation by X-ray
gives typically twice the optical estimation, being probably 
the maximal assessment of $\sigma_{\rm 1D}^2$.
For spiral galaxies, we interpret the asymptotic rotational velocity
into the velocity dispersion by the relation 
$\sigma_{\rm 1D}=v_{\rm rot}/\sqrt{2}$.

The basic data on the luminosities and the velocity dispersions
are taken from 
the table of Faber et al. (1989) 
for 332 ellipticals. The galaxies are selected from the original
sample on the condition that the surface brightness, angular size, and
velocity dispersion are measured. 
The X-ray data for 12 ellipticals are adopted from the table 
in Matsumoto et al. (1997).
Further, for 468 spiral galaxies, we combine the
I-band luminosities from Mathewson et al. (1996) 
and the asymptotic rotation velocities 
from Persic \& Salucci (1995).
We pick up the galaxies whose asymptotic velocities [$V_{as}$ as in
Persic \& Salucci (1995)] are given. Thus the
number of galaxies is smaller than the original one.

\section{Theory versus Observations}
\label{THOB}
\begin{figure}
\begin{center}
\psbox[width=150mm,vscale=1.0]{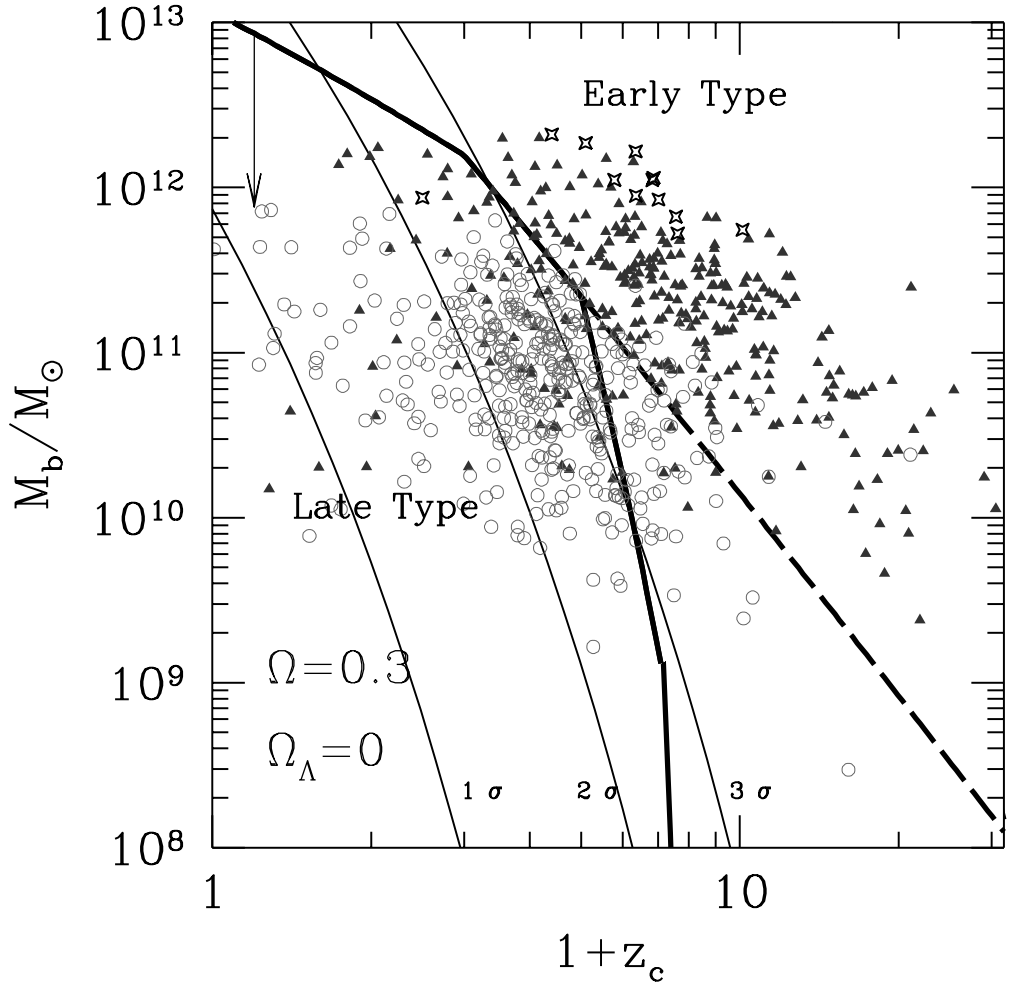}
\end{center}
\caption[f2.eps]{Comparison of observed galaxies with the bifurcation theory.
Small open circles represent spiral galaxies in Persic \& Salucci (1995). 
Small solid triangles denote the E and E-S0 galaxies in Faber et al. (1989). 
Open stars are elliptical galaxies from  X-ray observations in Matsumoto
et al. (1997). The thick solid and dashed lines represent the
 bifurcation mass $M_{\rm SB}^{\rm max}$, respectively for model A and
 B, where both are identical at $z_{\rm c}<4 $.  Three thin solid lines
 marked as $1\sigma$, $2\sigma$, and $3\sigma$ are the prediction in the
 CDM cosmology, where $\sigma$ is the variance of CDM perturbations. The
 downarrow in the upper-left corner of the panel shows the maximal shift
 of the bifurcation mass from $M_{\rm SB}^{\rm max}$ to 
$M_{\rm SB}^{\rm min}$ when including dark matter.}
\label{fig2}
\end{figure}

%Coincidence
In Figure \ref{fig2}, the bifurcation theory is confronted with 
observations.
In this figure, cosmological parameters are assumed as $\Omega=0.3$,
$\Omega_\Lambda=0$, 
$h=0.7$, and $\Omega_{\rm b} h^2=0.02$. All of these
values are plausible in the light of the recent observations,
although the value of $\Lambda$ parameter is still controversial. 
The small filled triangles and open circles represent respectively 
the observed elliptical and spiral galaxies. 
The open stars are the X-ray luminous elliptical galaxies.
We find that the two types of galaxies are successfully divided 
by the theoretical bifurcation mass scale.
This implies that the self-shielding against 
UV background radiation is practically related to 
the origin of the galactic morphology.
The difference between model A and B is not so significant, although
model A predicts a bit more ellipticals than observed.
Also, it is worth noting that few spiral galaxies reside above
the bifurcation mass scale, 
whereas there are a noticeable number of elliptical galaxies
which seem to have formed at lower redshift epochs 
than predicted by the present theory.
Since a condition for constant velocity dispersion gives a relation
$M \propto (1+z)^{-3/2}$ in this diagram,
most of discrepant low-redshift ellipticals turn out to be 
relatively luminous galaxies with higher stellar velocity dispersions.
Intriguingly, Gonzalez (1993) has reported,
using strengths of $H_\beta$ and [MgFe] indices, the evidence
for intermediate-age populations in elliptical galaxies. 
In addition, it has been argued that the galaxy merger with
burst-like star formation can lead to the formation of
elliptical galaxies (\cite{LT78}; Barnes \& Hernquist 1991, 1996; 
\cite{KC98}). 
Also, Shier \& Fischer (1998) suggest, by studying stellar kinematics of
starbursting infrared-luminous galaxies with obvious morphological
signatures of merger, that they can be progenitors of elliptical galaxies 
with high stellar velocity dispersions. 
Thus, the discrepant low-redshift ellipticals 
might not be of primordial collapse, but could be a category of
merger remnants.

In Figure \ref{fig2}, $1\sigma$, $2\sigma$, and $3\sigma$
CDM density perturbations normalized by COBE data 
(\cite{HuS96}; \cite{BW97}) are also plotted.
We find that elliptical galaxies form typically 
from $\sim 4\sigma$ perturbations 
and spirals do from $\sim 2\sigma$ peaks. 
If we change the cosmology, the results alter to some degree, 
because the bifurcation mass, the CDM spectra and the data points
of elliptical, and spiral galaxies are relatively changed. 
The main change of Figure \ref{fig2} comes from changing $\Omega$.
The collapse epoch $z_c$ for the observational data is dependent upon 
$\Omega$ as roughly $(1+z_c) \propto \Omega^{-0.74}$  
for $\Omega_\Lambda=0$ universe.
This dependence is mainly due to the evaluation of dark mass in a
galaxy. For instance, if we employ $\Omega=1$ (although it is unlikely), 
$1+z_{\rm c}$ becomes almost 2.5 times smaller than 
the estimates in Figure \ref{fig2}.
In this case, the theory and the observational data seem perceptibly 
discrepant with each other.
On the other hand, Figure \ref{fig2} is insensitive to 
the cosmological constant parameter $\Omega_\Lambda$.
In fact, even if we employ $\Omega=0.3$ and $\Omega_\Lambda=0.7$,
the relative position between the bifurcation mass and
observations remains almost unchanged, although
the correspondence between CDM fluctuations and observations
shifts from $\sim 4\sigma$ to $\sim 3\sigma$ for ellipticals.
In this case, the theory is in a good agreement with the observations.
Further details of 
the dependences on the cosmological model will be discussed in a
forthcoming paper.

%density-morphology relation 
The present results are also intriguing from a view point of
the statistics of galaxies.
Bardeen et al. (1986) have shown that 
higher $\sigma$ peaks reside preferentially in denser regions 
rather than in low-dense regions. 
As a result, the so-called density morphology relation 
can be explained as a natural consequence of the bifurcation theory.
Furthermore, it is known that the specific angular momenta $J/M$ of spirals 
are systematically greater by a factor of $\gsim 3$ than $J/M$ of 
ellipticals in the same mass range (\cite{Fal83}).
Based upon the tidal origin of the angular momentum, 
we expect $J/M \propto (1+z_c)^{-1/2}\nu^{-1}$, 
where $\nu \equiv \delta/\sigma$ (\cite{HP88}).
If ellipticals form from $\sim 4\sigma$ and
spirals do from $\sim 2\sigma$, then we anticipate
$(J/M)_{\rm spiral}\approx 3 (J/M)_{\rm elliptical}$.
\section*{Acknowledgments}
We appreciate the comments by the anonymous referee 
which helped us to improve the paper.
We thank M. Nagashima who kindly provided the data of mass-to-light
ratio. We also thank Y. Kanya and R. Nishi for useful information 
and comments. The analysis has been made with computational facilities 
at the Center for Computational Physics in University of Tsukuba. 
This work is supported in part by Research Fellowships of the Japan 
Society for the Promotion of Science for Young Scientists, No. 2370 (HS)
and by the Grants-in Aid of the Ministry of Education, Science, 
Culture, and Sport, 09874055 (MU). 

%%%%%%%%%%%%%%%%%%
% (4) Appendices %
%%%%%%%%%%%%%%%%%%
% NO APPENDICES

%%%%%%%%%%%%%%%%%
% (5)References %
%%%%%%%%%%%%%%%%%

\newpage
%%%%%%%%%%%%%%%%%%%%%%%
% (6) Figure Captions %
%%%%%%%%%%%%%%%%%%%%%%%
\onecolumn

\end{document}